\documentclass[10pt,preprint]{aastex}
\usepackage{epsfig}
\begin{document}
\newcommand{\PSbox}[3]{\mbox{\includegraphics{#1}\hspace{#2}\rule{0pt}{#3}}}
\newcommand{\degree}{\ensuremath{^\circ}}
\newcommand{\be}{\begin{equation}}
\newcommand{\ee}{\end{equation}}
\newcommand{\ein}{e_\mathrm{in}}

\title{Observational biases in determining extrasolar planet eccentricities in single-planet systems}

\author{Nadia L. Zakamska\altaffilmark{1,2}, Margaret Pan\altaffilmark{2,3}, Eric B. Ford\altaffilmark{4,3}
\altaffiltext{1}{{\it Spitzer} fellow, John N. Bahcall fellow}
\altaffiltext{2}{School of Natural Sciences, Institute for Advanced Study, Princeton NJ 08540}
\altaffiltext{3}{Kavli Institute for Theoretical Physics, University of California, Santa Barbara, CA 93106}
\altaffiltext{4}{Department of Astronomy, University of Florida, 211 Bryant Space Science Center, PO Box 112055, Gainesville, FL, 32611-2055}
}

%\date{\today}

\begin{abstract}
We investigate potential biases in the measurements of exoplanet orbital parameters obtained from radial velocity observations for single-planet systems. We create a mock catalog of radial velocity data, choosing input planet masses, periods, and observing patterns from actual radial velocity surveys and varying input eccentricities. We apply Markov Chain Monte Carlo (MCMC) simulations and compare the resulting orbital parameters to the input values. We find that a combination of the effective signal-to-noise ratio of the data, the maximal gap in phase coverage, and the total number of periods covered by observations is a good predictor of the quality of derived orbit parameters. As eccentricity is positive definite, we find that eccentricities of planets on nearly circular orbits are preferentially overestimated, with typical bias of $1-2$ times the median eccentricity uncertainty in a survey (e.g., 0.04 in the \citealt{butl06} catalog). When performing population analysis, we recommend using the mode of the marginalized posterior eccentricity distribution to minimize potential biases. While the \citet{butl06} catalog reports eccentricities below 0.05 for just 17\% of single-planet systems, we estimate that the true fraction of $e\le 0.05$ orbits is about $f_{0.05}=38\pm 9$\%. For planets with $P>10$ days, we find $f_{0.05}=28\pm 8$\% versus 10\% from \citet{butl06}. These planets either never acquired a large eccentricity or were circularized following any significant eccentricity excitation.
\end{abstract}

\keywords{methods: statistical -- planetary systems -- techniques: radial velocities}

\section{Introduction}

\begin{figure}
\epsscale{0.7}
%\plotone{picture_distribution_paper.eps}
\plotone{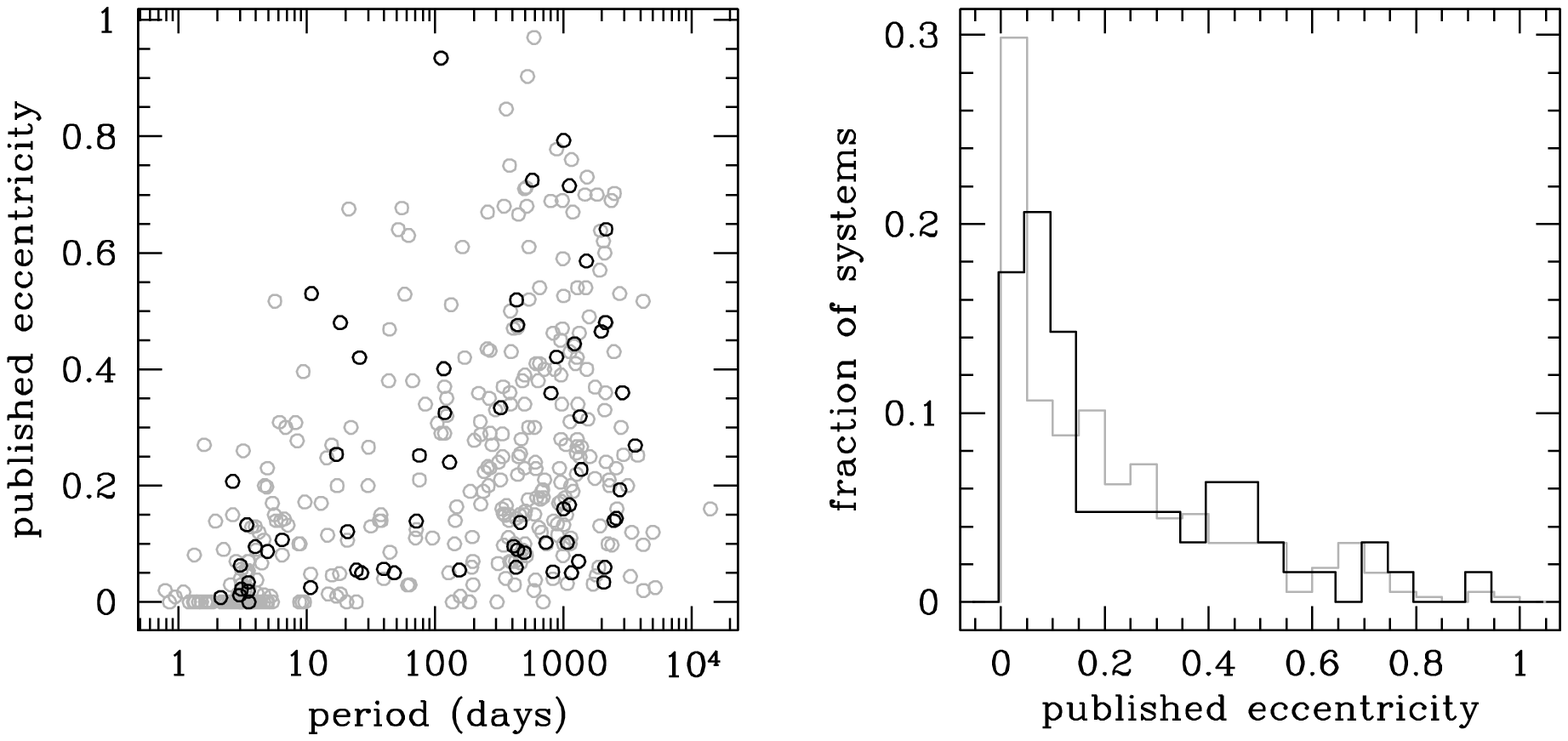}
\figcaption{Left: periods and eccentricities of all known radial velocity planets (grey) and 63 systems from B06 planets for which a single planet provides a good fit, as described in \S\ref{sec_compare} (black). Right: comparison of eccentricity distributions. Values of periods and eccentricities of 385 radial velocity planets were taken from exoplanet.eu as of end April 2010 (the list includes short-period planets discovered using the transit method but which have radial velocity observations). The two eccentricity distributions are consistent with each other in the sense of the Kolmogorov-Smirnov test. The fraction of planets with published eccentricities $< 0.05$ is $f_{0.05}=17-30\%$ (the two values are for B06 and for exoplanet.eu planets, respectively), and the fraction of those with $e\le 0.1$ is $f_{0.1}=38-41\%$. Excluding planets with $P<4$ days, $f_{0.05}=9-17\%$. \label{pic_1}}
\end{figure}

One of the most surprising properties of the $\sim$300 extrasolar planets found by radial velocity surveys is that their orbital eccentricities are much higher than those of the planets in the Solar System (Figure~\ref{pic_1}). Indeed, after excluding planets with short orbit periods ($P\la 4$ days) that have likely been influenced by tidal circularization, only $\sim$17\% of these planets have eccentricities $\la 0.05$\footnote{This value is obtained by taking eccentricities reported on exoplanet.eu (as of end April 2010) for radial velocity planets and discarding those with $P\le 4$ days.}. Several groups have proposed mechanisms able to excite extrasolar planet eccentricities to the levels found in radial velocity surveys \citep{trem04}. Some examples are planet scattering (e.g., \citealt{rasi96, weid96}), perturbations from a wide binary companion (e.g., \citealt{holm97}), and perturbations from passing stars (e.g., \citealt{laug98, zaka04, malm07}).

Subsequent studies of planets in binary systems \citep{marz05, take05, fabr07}, close stellar encounters \citep{malm09} and planet-planet scattering \citep{ford01, marz02, adam03, chat08, ford08a, juri08} have made detailed predictions for the distributions of extrasolar planet orbit properties. These studies typically use the distribution of the published best-fit orbit eccentricities as a proxy for the true underlying eccentricity distribution. They are therefore vulnerable to any significant differences between the measured and intrinsic distributions. To mitigate the effects of uncertainties in the distribution --- and in deviation from standard practice --- \citet{ford08a} performed Bayesian analyses of single planet systems, so that they could account for the uncertainty in the measurement of orbit parameters when comparing the predicted distribution to observations. While this approach has the advantage of emphasizing well-measured planets and de-emphasizing poorly measured ones, it still may be affected by biases. For example, since eccentricity is positive definite, an eccentricity measurement for a planet on a circular orbit can only overestimate the eccentricity.  Current observational estimates of the eccentricity, or of the posterior distributions for the eccentricity, are likely biased towards larger eccentricities for nearly circular planets \citep{lucy71, shen08, otoo08}.

There are several reasons to be interested in accurate measurements of extrasolar planet eccentricities, especially at small eccentricities where measurement biases are most significant. For example, precise determination of eccentricity distribution is important for short-period planets which can be circularized by tidal interactions with the star. By examining which of the planets with short tidal times have fully circularized and which retain a significant eccentricity, one can calibrate models of tidal dissipation and identify recent or ongoing eccentricity excitation episodes \citep{ford06b, mats08, naga08, baty09}. For planets beyond the reach of tidal circularization, an accurate eccentricity distribution, and in particular the fraction of low-eccentricity planets, may provide a probe of planet formation processes. For example, current data suggest that planet-planet scattering models predict fewer planets on nearly circular orbits than are observed \citep{juri08,chat08}.

In this paper we quantify the biases of extrasolar planet eccentricities when measured from radial velocity observations. We create a mock catalog of radial velocity data, choosing planet masses, orbit periods and observing patterns to mimic those of actual radial velocity surveys (\S\ref{sec_mock}). Using a Bayesian framework and Markov chain Monte Carlo (MCMC) simulations, we calculate the posterior probability distribution for each mock data set, along with several summary statistics for each data set and each orbit parameter (\S\ref{sec_methods}). We examine several different eccentricity estimators and make recommendations for minimizing the effects of bias. More generally, we investigate the quality of orbit parameter determination in modern radial velocity surveys and investigate the reliability of determining orbit parameters as a function of the quality of data sets in \S\ref{sec_quality}. Based on these results, we estimate the underlying distribution for the eccentricities of extrasolar planets in \S\ref{sec_true}. We summarize our results in \S\ref{sec_summary}.

\section{Experimental design}
\label{sec_mock}

Our investigation is simple in concept: we input a set of realistic orbit parameters for an extrasolar planet, generate a radial velocity curve for its host star, compute the orbit parameters from the radial velocity curve in a blind experiment and compare output with input. However, the many choices necessary to produce a realistic radial velocity curve make setting up such an experiment quite a challenge in itself, especially if we aim to provide a prescription for correcting the biases in real surveys. For example, generating radial velocity curves from scratch requires not only assumed intrinsic distributions of planet and host star properties but also a way to account for the varying sensitivities, observing strategies and target selection methods of different surveys. We sidestep most of these challenges by taking our input orbit parameters, observational errors and time sampling directly from the real systems in the catalog of Butler et al. (2006; hereafter B06). These authors present orbital solutions for 172 exoplanets within 200 pc of the Sun and provide public radial velocity data sets for those from the California Carnegie Planet Search. It is the largest refereed catalog of exoplanets discovered via radial velocity observations and it contains most of the systems for which radial velocity data are publicly available. Our approach is then to investigate the possible biases in determination of orbit parameters in this particular catalog.

To create mock planet systems, we take the best-fit period $P$, velocity amplitude $K$, argument of periastron $\omega$ and time of passage of the periastron $t_p$ of the dominant planet in each of the 91 systems in the B06 catalog with fewer than 90 observations (the others take prohibitively long to analyze in large numbers), and use times of observations and observational errors from the corresponding radial velocity curves. Among these 91 systems, the median data set contains 40 data points, covers 8 planet periods and has a radial velocity uncertainty of 3.3 m/s. For each system we generate synthetic radial velocity data sets $v(t)$ for five input eccentricities $e=$~0, 0.05, 0.1, 0.3, 0.6 using the Keplerian model \citep{murr99}
\begin{equation}
v(t)=K\left[\cos(\omega+T(t))+e\sin(\omega)\right],\label{eq_kepler1}
\end{equation}
where $T$ is the true anomaly. The argument of periastron is measured from the line in the orbit plane where it intersects the sky plane and the planet is approaching the observer. At each actual observation time $t_i$, we generate a simulated velocity ($v_i$) as a random variable normally distributed about a ``true'' value of $v(t_i) + C$ with dispersion $\sqrt{\sigma_{{\rm obs},i}^2+\sigma_J^2}$. Here $C$ is the constant systemic velocity of the system, $\sigma_{{\rm obs},i}$ are measurement uncertainties taken directly from the real data sets and $\sigma_J=3.5$~m/s is a fixed velocity ``jitter'' adopted to account for astrophysical noise such as stellar photospheric activity \citep{wrig05}.

Each resulting mock radial velocity data set consists of the times of observation $t_i$, mock radial velocity measurements $v_i$, and errors of measurement $\sigma_{{\rm obs},i}$. For each input eccentricity, we create five realizations of the radial velocity data with a different set of Gaussian random variables for each realization, using $\sqrt{\sigma_{{\rm obs},i}^2+\sigma_J^2}$ as the width of the Gaussian distribution. We then perform Bayesian analyses of the resulting 2275 ($=91$ systems $\times$ 5 eccentricities $\times$ 5 realizations) mock radial velocity data sets.
   
\section{Analysis of mock and observational data}
\label{sec_methods}

The Bayesian analysis of radial velocity observations is described by \citet{ford05, greg05, ford06a, greg07,ford07, bala09}, among others. In this Section, we review the essential elements and point out the details that differ from those in previous work.

\subsection{Priors and likelihood}
\label{sec_priors}

To ensure an efficient performance of the global search algorithm (\S\ref{sec_global_search}), we adopt simple, analytic, separable priors that provide a reasonable first approximation to the exoplanet distribution discovered by radial velocity planet searches. They take the form
\begin{equation}
p_{\rm all}(P,K,e,\omega,M_0,C,\sigma_J) = p_P(P) p_K(K) p_e(e) p_{\omega}(\omega) p_M(M_0) p_C(C) p_{\sigma_J}(\sigma_J),
\end{equation}
where $M_0$ is chosen to be the mean anomaly at the middle of the time series. The prior for the orbit period is uniform in logarithm of the period: $p_P(P) = P^{-1} / \ln(P_{\max}/P_{\min})$. For the global search (\S\ref{sec_global_search}) we impose hard limits, $P_{\min} = 2$ days and $P_{\max}= \pi (t_{\rm max}-t_{\rm min})$, where $t_{\rm max}-t_{\rm min}$ is the time interval between the first and the last observations. These limits are chosen to be far away from the solution for any planet clearly detected from the present data \citep{ford06a}. The priors for the radial velocity amplitude and jitter are modified Jeffreys priors \citep{greg05} of the form $p_x(x; x_0, x_{\max}) = (1+x/x_0)^{-1} / \ln(1+x_{\max}/x_0)$, where $x$ is either $K$ or $\sigma_J$. The hard upper limit for the amplitude, $K_{\max} = (P/P_{\min})^{-1/3} / \sqrt{1-e^2}\times 1690$~m/s, corresponds to the radial velocity amplitude produced by a $\sim 10 M_{\rm Jup}$ planet orbiting a solar mass star in the plane containing the line of sight. The upper limit for the jitter, $\sigma_{J,\max} = 1690$~m/s, is based on the same amplitude for a circular planet with period $P_{\min}$. The scale parameters ($K_0$ and $\sigma_{J,0}$) are set to 0.1 m/s so as to prevent a peak in the posterior at small amplitudes due to a divergence at zero in a Jeffreys prior. The prior for the eccentricity is uniform between zero and unity, and we further discuss the effects of this assumption in \S\ref{sec_2d}. The priors for the argument of the periastron and for the mean anomaly are uniform between 0 and $2\pi$.  The prior for the velocity offset is uniform and not bounded. 

We assume that each radial velocity observation is independent and normally distributed about the true value with a variance $\sigma_{\rm obs,i}^2+\sigma_{J}^2$.  Therefore, the likelihood is given by
\begin{equation}
L({{\bf d} | \mbox{\boldmath$\theta$}} )
 = (2\pi)^{-N/2} 
   \left[\prod_{i=1}^{N} (\sigma_{{\rm obs},i}^2+\sigma_{J}^2)
   \right]^{-1/2} 
   \exp (-\chi^2(\mbox{\boldmath$\theta$})/2),
\end{equation}
and
\begin{equation}
\chi^2(\mbox{\boldmath$\theta$} ) = \sum_{i=1}^{N} \frac{(v_i - v(t_i))^2}{\sigma_{{\rm obs},i}^2+\sigma_{J}^2}.
\end{equation}
The posterior probability distribution is given by Bayes' theorem,
\begin{equation}
p(\mbox{\boldmath$\theta$} | {\bf d})
 = p(\mbox{\boldmath$\theta$}) L( {\bf d} | \mbox{\boldmath$\theta$}) 
   / p({\bf d}),
\end{equation}
where $p({\bf d}) = \int p(\mbox{\boldmath$\theta$}) L({\bf d}|\mbox{\boldmath$\theta$})\, d\mbox{\boldmath$\theta$}$ is a normalizing constant that need not be computed for our purposes of parameter estimation within a single model.

We perform our Bayesian analysis in two steps.  First, we perform a global search to identify the dominant mode(s) of the posterior distribution (\S\ref{sec_global_search}). The output of this global search step is used to generate the initial states of the Markov chains for a MCMC analysis (\S\ref{sec_mcmc}).

\subsection{Global Search}
\label{sec_global_search}

In the global search we perform a brute force integration over the variables $\left\{P, e, M_0\right\}$ using two quasi-random number generators (QRNGs), perform a one-dimensional numerical integration over $\sigma_J$, and expand the arguments of the exponents using Taylor series in the remaining integrals (the Laplace approximation; \citealt{cumm04, ford08b}).  For the outer loop, the first QRNG uses a one-dimensional Sobol sequence to generate the orbit frequency $1/P$ for periods between $P_{\min} = 2$ days and $P_{\max} = \pi (t_{\rm max}-t_{\rm min})$.  We evaluate the (unnormalized) posterior (i.e., prior times likelihood) at each of $N_{\rm per}$ periods. Following \citet{greg07}, 
\begin{equation}
N_{\rm per} = \max\left\{10^5, \min\left\{ 10^4,
    2\,\mathrm{ceil}\left[P_{\max}/P_{\min}-1\right]
    \mathrm{ceil}[1+1.6 \times ({\rm S/N}) \times (t_{\rm max}-t_{\rm min})/(P_{\max}\sqrt{N})] \right\} \right\},
\end{equation}
where ${\rm S/N} = N^{-1/2} \left[ \sum_{i=1}^{N} (v_i-C_{\mathrm{bf}})^2/\sigma_{{\rm obs},i}^2 \right]^{1/2}$ is an estimate of the combined signal to noise and $C_{\mathrm{bf}}$ is the best-fit constant velocity for the given data set.

At each sampled orbit period, the second QRNG uses a two-dimensional Sobol sequence to generate the eccentricity $e$ and mean anomaly at epoch $M_0$ uniformly over their full range of eccentricities and periastron angles.  This Sobol sequence contains 256 to 4,096 pairs of $e$ and $M_0$.  The number of samples is chosen so that the accuracy is better than 10\% (20\%) depending on whether the period being considered contributes more than (less than) one part in $10^8$ of the current estimate for the posterior probability marginalized over all periods.  The accuracy of the (unnormalized) marginalized posterior probability for a given period is estimated based on the first and second half of the Sobol sequence after 256, 512, 1024, 2048, and 4096 samples.

For each set $(P, e, M_0)$, we integrate numerically over $\sigma_J$ using the adaptive Gauss-Kronrod integration method implemented in GNU Science Library. For each set $(P, e, M_0, \sigma_J)$, we estimate the (unnormalized) posterior probability marginalized over all the remaining ``linear'' model parameters $K$, $\omega$, $C$ using the Laplace method.  By changing variables from $K$ and $\omega$ to $K\cos \omega$ and $K\sin \omega$, we can write the radial velocity model as a linear function of the remaining parameters. For fixed values of $(P, e, M_0, \sigma_J)$, there is then a single global minimum of $\chi^2$ which can be efficiently solved for via matrix algebra \citep{wrig09}. We use singular value decomposition to solve for the best-fit values of the linear model parameters. We then approximate the (unnormalized) marginalized posterior probability based on the value of the prior, the likelihood, the determinant of the inverse covariance matrix, and a Jacobian all evaluated at the best-fit value.  The Jacobian, $J = 1/\left|K\right|$, is needed since the Laplace approximation depends on the model parametrization.  Our model is linear in the variables $K\cos\omega$ and $K\sin\omega$, but we use priors that are uniform in $K$ and $\omega$.

After integrating over all model parameters, we identify the dominant mode of the posterior probability distribution and draw samples of the model parameters from this mode for use as initial states in the Markov chain Monte Carlo analysis described in \S\ref{sec_mcmc}. The above global search algorithm is parallelized using OpenMP and is applied separately to each data set considered. In principle, the above global search algorithm is fully automated and does not require human intervention to provide knowledge about the location of the best-fit model.  In practice, some data sets failed initially and were rerun with $N_{\rm per}$ larger by a factor of 4.

\subsection{MCMC analysis}
\label{sec_mcmc}

For each data set, we perform a MCMC analysis closely following the methods of \citet{ford06a}.  All our Markov chains take steps in the variables $1/P$, $K$, $e$, $e\sin(\omega+M_0)$, $e\cos(\omega+M_0)$, $\omega+M_0$, and $\ln \sigma_J$.  The velocity offset is explored using Gibbs sampling. However, when two model parameters are strongly correlated, steps in the two variables individually can be very inefficient.  Much larger steps are possible if they are taken in sets of variables that have weaker correlation. \citet{ford06a} identified expanded sets of stepping variables that greatly accelerate the convergence of Markov chains for a variety of types of systems and radial velocity data sets. We use this expanded set of stepping variables to help our Markov chains converge more rapidly. Therefore, for data sets where the best-fit solution (identified from the above global search) has a period of greater than 10 days, we also allow our Markov chains to take steps in the variables $\omega$, $K\cos\omega$, $K\sin\omega$, $\omega+T_0$, $t_p$, and $\ln K\sqrt{1-e}$, where $T_0$ is the true anomaly at epoch. 

Before calculating the Markov chains to be used for inference, we first calculate chains with variable step scale factors so as to determine step scales that result in efficient exploration of parameter space. In the vast majority of cases, our standard initial guesses for the step sizes resulted in convergence on step scale factors that gave acceptance rates of $\sim40\%$; in about 5\% of cases we had to manually adjust the initial guess for the step size for the algorithm to identify usable step scale factors. We then discard these initial chains and produce Markov chains with fixed step scale factors for parameter estimation.

For each data set, we compute five Markov chains consisting of at least $10^6$ states each which we use to test for any signs of non-convergence as described below.  Once a set of Markov chains is accepted, we draw a random sample of $10^4$ states from the final 80\% of all five chains.  This sample is used for inference such as calculating summary statistics as described in \S\ref{sec_statistics}.

To flag non-convergence we calculate the Gelman-Rubin test statistic and estimate the correlation length for each for the variables listed in \S\ref{sec_priors} (regardless of the orbit period), as well as $M_0$, as described in \citet{ford06a}.  As a practical matter, our Markov chain calculations use a prior uniform in $\ln P$ without any bounds.  Therefore, we require that the chain converge upon a plausible range of orbit periods with significant weight between $P_{\min}$ and $P_{\max}$.  In several cases, much longer chains were needed in order to pass the above convergence tests.

Of the 2275 mock data sets, we exclude 17 which failed convergence tests despite attempts to modify step size and direct the chain toward a different global search solution. We exclude an additional 31 chains which nominally converged but whose final periods are too poorly determined ($\sigma[P]/P>0.5$, where $\sigma[P]$ is defined in detail in the next section). We further exclude another 31 chains whose output period deviates too much from the input period, that is, those with $\log_{10}|P_\mathrm{out}/P_\mathrm{in}|>0.3$. Reasons for the poor performance of most of the 79 chains thus rejected (3.5\% of the total number investigated) are easily identified (e.g. too few data points, low signal from the planet, or an orbit period commensurate with yearly gaps in observing coverage).

\section{Determination of orbit parameters}
\label{sec_quality}

To interpret our MCMC results, we calculate from each multi-dimensional posterior sample a small set of summary statistics for each orbit parameter. To do this, we marginalize each chain over all orbit parameters except one to estimate the one-dimensional posterior probability distribution for this one parameter. We then find summary statistics to describe the one-dimensional distributions. The summary statistics typically reported in planet discovery papers are the best-fit value of a given orbit parameter and an uncertainty often interpreted as a 1-sigma confidence interval with the model of a Gaussian distribution in mind. However, for an arbitrary one-dimensional marginalized posterior distribution, care must be taken to choose parameters that accurately describe the sample. In \S\ref{sec_statistics} and \S\ref{sec_2d} we consider several options.

\subsection{Output quality for one-dimensional distributions}
\label{sec_statistics}

As candidates for the best-fit value of an orbit parameter we compare the median, mean, and mode of the one-dimensional marginalized posterior distribution. We estimate the mode --- the location of the maximum probability density --- by locating the densest clump of values in the $10^4$-state subsample from the output Markov chain. We proceed by locating the smallest contiguous interval containing 9999 of the 10000 sample values, then that containing 9998, then 9997, and so on. When the smallest contiguous interval containing $k-1$ sample values shifts away from the position of the smallest interval containing $k$ samples by more than $0.2k$ samples, we assume that we have reached the scale of density irregularities present due to finite sampling of the probability distribution. We therefore take the interval containing $k$ values as representative of the position of the densest clump of samples and take the median of the $k$ values --- the center of the clump --- as an estimate of the mode\footnote{The IDL code with our mode estimator is provided as an online supplement to this paper.}.

\begin{figure}
\epsscale{0.7}
%\plotone{picture_nchain_1d_log.eps}
\plotone{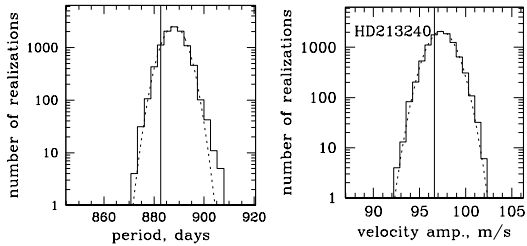}
\figcaption{Example marginalized posterior parameter distributions obtained from our MCMC simulations for a typical mock radial velocity data sets. The input eccentricity was 0 and the other orbit parameters were from HD213240b. The left panels shows the marginalized period distribution and the right panel shows the velocity amplitude distribution. The vertical line shows the input value. The dashed line shows a Gaussian distribution with the dispersion determined from the 68\% credible interval. \label{pic_1d}}
\end{figure}

We define a 68\% credible interval as the smallest contiguous interval containing 68\% of all points in the chain. For a Gaussian distribution, this percentage corresponds to the fraction of the distribution within 0.994458 standard deviations of the mean; by analogy, we define the measurement precision as $1/(2\cdot 0.994458)$ times the credible interval. Figure~\ref{pic_1d} shows example marginalized posterior distributions for periods and velocity amplitudes. The comparison between our observed distributions and a Gaussian function can be parametrized by comparing the 68\% credible intervals to analogously defined 95\% and 99\% ones. The median relationships between these values suggest that most of our posterior distributions for period and velocity are close to Gaussian, with $\sigma_{99}$ only about 6\% above $\sigma_{68}$. Therefore, the non-Gaussianity of our posterior distributions is not nearly as severe as that found by O'Toole et al. (2008), who used $\chi^2$ minimization to obtain orbit parameters. These authors find a factor of 5 to 10 difference between values of $\sigma$ obtained from 68\% and 99\% credible intervals. Such deviations from Gaussianity are seen in only 9\% of our systems. None of our conclusions are affected by the difference between $\sigma_{68}$, $\sigma_{95}$ and $\sigma_{99}$, so hereafter we use ``precision'' to mean $\sigma$ defined on the basis of the 68\% credible interval. We indicate the relevant orbit parameter in brackets, e.g., $\sigma[K]$ is the precision of the velocity amplitude measurement.

\subsection{Comparison of summary statistics for eccentricity}
\label{sec_2d}

\begin{figure}
\epsscale{0.7} 
%\plotone{picture_nchain_2d.eps} 
\plotone{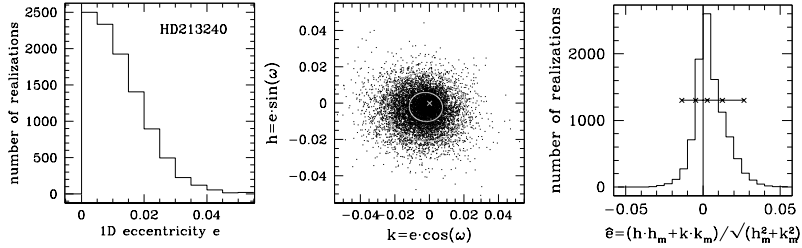}
\figcaption{The left panel shows the marginalized posterior eccentricity distribution for the same simulated radial velocity data set as in Figure~\ref{pic_1d}. The input eccentricity was 0. Since eccentricity is positive definite, all eccentricities within the credible interval are positive. The middle panel shows projection of the Markov chain onto the $h-k$ plane. The grey cross marks the input value (0,0) and the grey ellipses were computed using principal component analysis to approximate two-dimensional 68\% and 95\% credible contours. The right panel shows the distribution of the value $\hat{e}$ derived from $h$ and $k$ ($h_{\rm m}$ and $k_{\rm m}$ are the median $h$ and $k$). Although the best-fit $\hat{e}$ is positive definite, in general the values $\hat{e}$ can be positive or negative, so the credible intervals may be sensibly defined. The horizontal line with points shows the median of the distribution, as well as the 68\% and the 95\% credible intervals.\label{pic_2d}}
\end{figure}

The left panel of Figure~\ref{pic_2d} shows the marginalized posterior distribution for eccentricity for a system with input eccentricity 0 (the same as shown in Figure~\ref{pic_1d}). Clearly, the one-dimensional output eccentricity distribution is highly asymmetric for nearly circular orbits because eccentricity is positive definite: the best-fit output eccentricities are always positive, and so are the minimal eccentricities allowed by the credible intervals defined above. Of the three measures discussed in the previous section (the mode, the median and the mean), the mode of the eccentricity distribution is the closest to the true input value.

The middle panel demonstrates the behavior of the Markov chain marginalized over all parameters except $e$ and $\omega$. The chain is displayed in the plane of the ``two-dimensional eccentricity'' components $h = e\sin\omega$ and $k = e\cos\omega$ \citep{murr99}. In these variables a random-walk Markov chain can easily explore the parameter space near the input value $e = 0$, so we investigate eccentricity measures written in terms of $h$, $k$. A constant probability density in the $e$, $\omega$ space (corresponding to the flat priors we adopted for these variables) diverges as $1/\sqrt{h^2 + k^2}$ when the variables are changed to $h$ and $k$. This leads to a divergence $\propto |\ln (h,k)|$ at $h,k = 0$  in the marginalized 1D posterior distributions for $h$, $k$. These spikes render the modes of the $h$, $k$ distributions unusable as summary statistics. While the probability density for each variable diverges, its integral converges, so the mean and the median of the $h$, $k$ distributions are less sensitive to this problem; the corresponding eccentricity measures are $\hat{e}_\mathrm{med}= \sqrt{h^2_\mathrm{med} + k^2_\mathrm{med}}$ and $\hat{e}_\mathrm{mean}=\sqrt{h^2_\mathrm{mean} + k^2_\mathrm{mean}}$.

In an effort to compensate for the $h$-$k$ singularity's effects on summary statistics, we weight each point of the chain by its $e$ value (the Jacobian between $e$, $\omega$ and $h$, $k$ coordinates); values calculated with the use of weights are denoted with a `w'. This weighting gives an estimate of the posterior distribution for the same data using priors flat in $h$, $k$. We calculate the weighted median and weighted mean of the posterior distributions of $h$ and $k$, as well as the values $\hat{e}_\mathrm{w-med}=\sqrt{h^2_\mathrm{w-med} + k^2_\mathrm{w-med}}$ and $\hat{e}_\mathrm{w-mean}=\sqrt{h^2_\mathrm{w-mean} + k^2_\mathrm{w-mean}}$. 

To calculate precision measures for two-dimensional eccentricities, we draw a line in the $h$, $k$ plane through the points $(0,0)$ and $(h_\mathrm{med}, k_\mathrm{med})$ and project each sample from the Markov chain onto this line. The coordinates of the projected Markov chain points are $\hat{e} = (h\cdot h_\mathrm{med}+k\cdot k_\mathrm{med})/\hat{e}_\mathrm{med}$ and can be positive or negative. Using the one-dimensional distribution approach described in the previous section, we can compute credible intervals for $\hat{e}$. The right panel of Figure~\ref{pic_2d} shows an example $\hat{e}$ distribution and the corresponding credible intervals, with the advantage being that they correctly include $e = 0$ when it lies within the 68\% two-dimensional confidence region. Since $\sigma[\hat{e}]$ and other precision measures based on the 1D confidence intervals for $h$, $k$, $h_\mathrm{w}$, $k_\mathrm{w}$ differ by only a few percent, we adopt $\sigma[\hat{e}]$ for all future use.

\begin{figure}
\epsscale{0.7} 
%\plotone{picture_ecc_measures_paper.eps}
\plotone{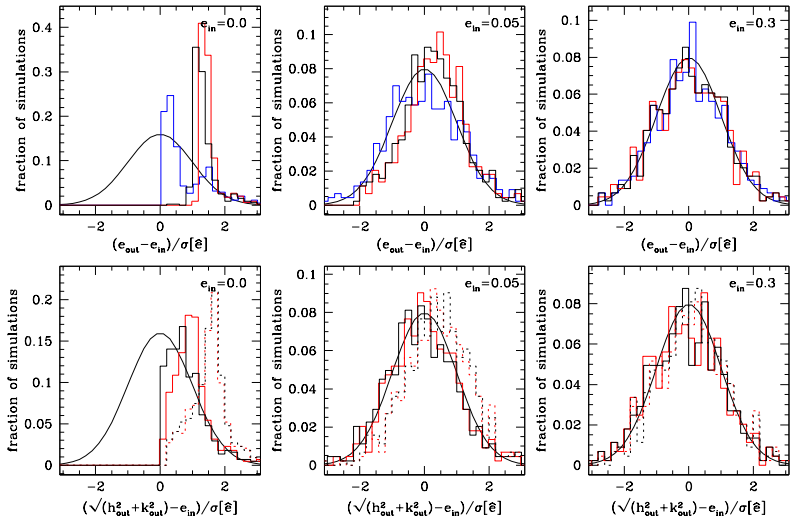}
\figcaption{Comparison of output eccentricity with input eccentricity for six eccentricity measures. Each panel shows histograms of the eccentricity bias ($\equiv e_{\rm out}-e_{\rm in}$) normalized to $\sigma[\hat{e}]$ together with a Gaussian distribution (smooth black line); the left, middle, and right columns show data for $\ein=0$,~0.05,~0.1 respectively. Top panels show data for eccentricity measures derived directly from the posterior distribution for eccentricity marginalized over all other parameters: the blue, red and black histograms correspond to the mode ($e_\mathrm{out}=e_\mathrm{mode}$), mean ($e_\mathrm{out}=e_\mathrm{mean}$), and median ($e_\mathrm{out}=e_\mathrm{med}$) of that distribution. While the mean and median are quite biased for input eccentricities $\la 0.05$, the bias vanishes at higher eccentricities. Bottom panels show the bias for alternative eccentricity measures defined in the $h-k$ plane, with red histograms for $\sqrt{h_{\rm mean}^2+k_{\rm mean}^2}$ and black for $\sqrt{h_{\rm med}^2+k_{\rm med}^2}$. Dotted histograms are for the values weighted by eccentricity, an estimate of the results we would have obtained using priors flat in $h$, $k$. \label{pic_ecc_measures}}
\end{figure}

In Figure~\ref{pic_ecc_measures} we compare the statistical properties of the summary statistics for eccentricity discussed above. All of our summary statistics $e_\mathrm{mean}$, $e_\mathrm{med}$, $e_\mathrm{mode}$, $\hat{e}_\mathrm{mean}$, $\hat{e}_\mathrm{med}$, $\hat{e}_\mathrm{w-mean}$, and $\hat{e}_\mathrm{w-med}$ are positive definite, so there is always a positive bias when $\ein = 0$. However, the strength of the bias varies with the definition used: $e_\mathrm{mode}$ and $\hat{e}_\mathrm{med}$ have nearly Gaussian distributions (or half-Gaussian for $\ein = 0$) around $\ein $, while $e_\mathrm{mean}$ and $e_\mathrm{med}$ for the one-dimensional eccentricity (red and black histograms in the top row) are typically $1-2$ times $\sigma[\hat{e}]$ larger than $\ein $ for nearly circular orbits. The magnitude of the bias is determined by the typical eccentricity uncertainty of the survey, which is 0.04 (median value of $\sigma[e]$) for the B06 catalog. Therefore, as seen in Figure~\ref{pic_ecc_measures}, the bias is strong for $\ein \leq 0.05$ and negligible for $\ein \geq 0.1$.

\subsection{Relationship between input quality and output quality}
\label{sec_corr}

In this section we aim to develop a set of diagnostic criteria which allow us to evaluate the quality of any radial velocity data set for the purpose of orbit parameter determination. We consider period, velocity amplitude, and eccentricity and we gauge the quality of these extracted values with the following metrics:
\begin{itemize}
\item $\sigma[P]$, $\sigma[K]$, $\sigma[\hat{e}]$ are the period, velocity amplitude, and eccentricity precisions defined based on the relevant 68\% credible intervals as explained in \S\ref{sec_statistics}. 
\item $P-P_\mathrm{in}$, $K-K_\mathrm{in}$, $e-\ein $, where ``in'' denotes the input value, give the ``bias''. Output $K$ and $P$ are taken to be the medians of the respective marginalized posterior distributions, whereas $e_{\rm mode}$ is taken to represent output eccentricity.
\item $(P-P_\mathrm{in})/P$, $(K-K_\mathrm{in})/K$ give the ``normalized bias'' for $P$, $K$.
\end{itemize} 
We also considered ``reliability'', the standard deviation of the five output values of each of $P$, $K$, $e$ extracted from the five different realizations generated using the same input system, as an indicator of sensitivity to noise. We find that, statistically, reliability is indistinguishable from precision --- confirming that $\sigma[K]$, $\sigma[P]$, $\sigma[\hat{e}]$ are indeed accurate measures of orbit parameter uncertainty --- so we do not discuss reliability further.

Similarly, for each data set we use several input quality metrics: 
\begin{itemize}
\item $N$ is the number of points in the data set.
\item $N_{\rm per}$ is the number of periods covered by observations, $(t_{\rm max}-t_{\rm min})/P$.
\item If $v_i$, $\sigma_{{\rm obs},i}$ are the observed radial velocities and their observational uncertainties, then the effective error-weighted precision of the data set is $\sigma_{\rm obs}=\langle 1/\sigma^2_{{\rm obs},i}\rangle^{-1/2}$. 
\item The effective signal-to-noise ratio is $K \sqrt{N}/\sigma_{\rm obs}$.
\item $\Phi_{\rm max}$ is the maximum gap in phase coverage in the phase-folded data set (we define the phase to run from 0 to 1).
\end{itemize}
Not all of the data set quality parameters are independent. We evaluate a rank correlation coefficient for every pair of these parameters (10 pairs), as well as the corresponding probability that the two parameters in the pair are uncorrelated. Two correlations are present at the 99.9\% signficance level: $N_\mathrm{per}$ is positively correlated with the effective signal-to-noise and $N$ is negatively correlated with $\Phi_\mathrm{max}$. With these caveats in mind, we look for correlations between the data set parameters and the output metrics for all orbit parameters.

\begin{figure}
\includegraphics*[scale=.5,angle=90]{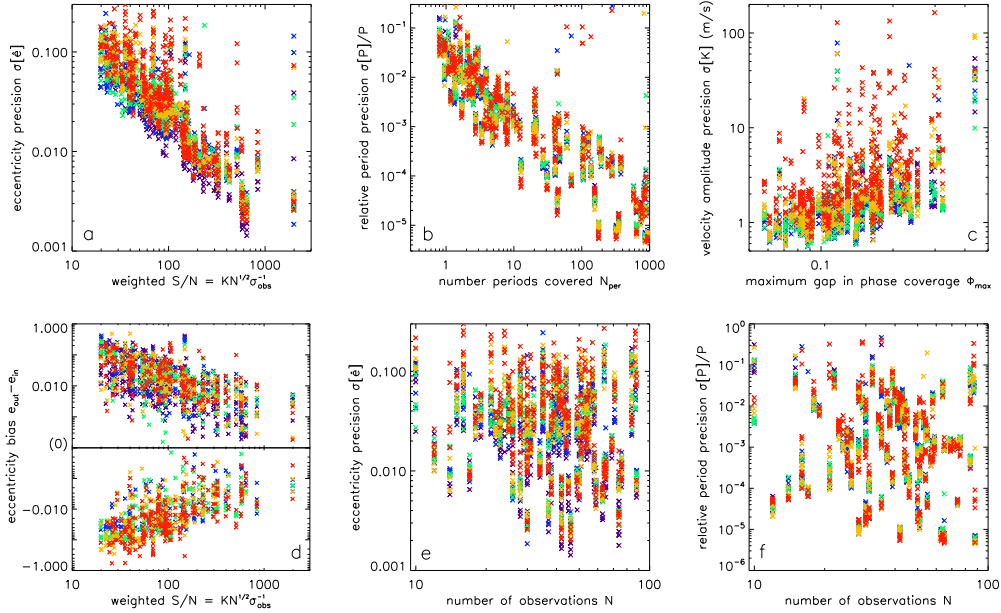} 
\figcaption{Scatter plots illustrating correlations between selected data set quality and output quality metrics: a) weighted signal-to-noise vs. eccentricity precision; b) number periods covered vs. normalized period precision; c) maximum gap in phase coverage vs. velocity amplitude precision; d) weighted signal-to-noise vs. eccentricity bias (the nine points corresponding to eccentricity biases of absolute value between $10^{-6}$ and $10^{-4}$ are not shown); e) number of observations vs. eccentricity precision; f) number of observations vs. normalized period precision. The wavelength of the color code increases with input eccentricity (purple for $\ein=0$, blue for $\ein=0.05$, green for $\ein=0.1$, orange for $\ein=0.3$, red for $\ein=0.6$). The pairs of input and output metrics in panels a), b), c) are those we found to be most strongly correlated. Guidelines for expected output quality from an RV data set of given input quality based on these strong correlations are discussed in \S\ref{sec_summary}. Panel d) clearly shows positive bias for small values of input eccentricity $e_\mathrm{in}=0$,~0.05,~0.1. Bias and uncertainty in eccentricity decline strongly with ${\rm S/N}_{\rm eff}$ (panels a, d), but are not correlated with $N$ (panels e, f). \label{pic_corr}} 
\end{figure}

The strongest correlations relate the eccentricity precision $\sigma[\hat{e}]$ and the effective signal-to-noise; the normalized period precision $\sigma[P]/P$ and the number of periods covered, $N_{\rm per}$; and the velocity amplitude precision $\sigma[K]$ and the maximum phase gap $\Phi_{\rm max}$ (Figure~\ref{pic_corr}). The corresponding Spearman's rank correlation coefficients are respectively $-0.844$, $-0.881$, $0.537$. To the extent we can generalize from our simulated data to other radial velocity data sets, these correlations suggest rough guidelines for the output precisions one can expect from a data set with given values of effective signal-to-noise, $\Phi_{\rm max}$, and $N_{\rm per}$ as follows. 
\begin{itemize}
\item An eccentricity precision better than $0.05$ requires an effective signal-to-noise greater than $\sim$40.
\item Observations over one complete orbital period typically result in a normalized period precision of 10\%, and a normalized period precision of 1\% requires a time baseline of $2-3$ periods.
\item Assuming radial velocity measurement uncertainties typical of the B06 catalog ($3.7\pm 1.8$ m/s), a velocity amplitude precision of 3~m/s requires a maximum phase gap of less than about $0.3$ orbit period for low- to moderate-eccentricity systems corresponding to our $e_\mathrm{in}=0$,~0.05,~0.1 and a maximum phase gap of less than about 0.15 orbit period for moderate- to high-eccentricity systems corresponding to our $e_\mathrm{in}=0.3$,~0.6.
\end{itemize}
These guidelines correspond to median relations between the input and output quality metrics. Upper envelopes of these relations can be parametrized as
\begin{eqnarray}
\log\sigma[\hat{e}]=0.48-0.89\times \log(K\sqrt{N}/\sigma_{\rm obs});\label{eq_cor1}\\
\log(\sigma[P]/P)=-1.23-1.00\times \log(N_{\rm per});\label{eq_cor2}\\
\log(\sigma[K], {\rm m/s})=1.36+0.89\times \log(\Phi_{\rm max}).\label{eq_cor3}
\end{eqnarray}
These relations are obtained by fitting power laws to the correlations in Figure~\ref{pic_corr} and then adjusting the normalization so that 90\% of all points are below relations (\ref{eq_cor1})-(\ref{eq_cor3}).
%We find that for some data sets, an eccentricity precision of 0.05 may %require an effective signal-to-noise of $100-200$; a normalized period %precision of 10\% may require observations over more than two orbit periods; %and/or a velocity amplitude precision of 3~m/s may require a maximum phase %gap of less than 0.1 orbit period for low- to moderate-eccentricity systems %or less than 0.07 orbit period for moderate to high eccentricities. These %numbers correspond to the upper envelope of the bulk of our mock systems as %shown in Figure~\ref{pic_corr}.

Notably, we see no correlation between our output quality metrics and the number of observations $N$ (Figure~\ref{pic_corr}e,f). This seems counter-intuitive and appears to contradict the findings of \citet{shen08}. The explanation for this apparent discrepancy is that our data involve an ensemble of systems rather than a single system viewed in a variety of observing situations. For an ensemble, the relation between output quality and $N$ is complicated because new planet systems are generally published when they have passed some minimum reliability threshold regardless of how many observations were made to attain that threshold.

\begin{figure}
\epsscale{0.7}
%\plotone{picture_paper_shen_compare.eps}
\plotone{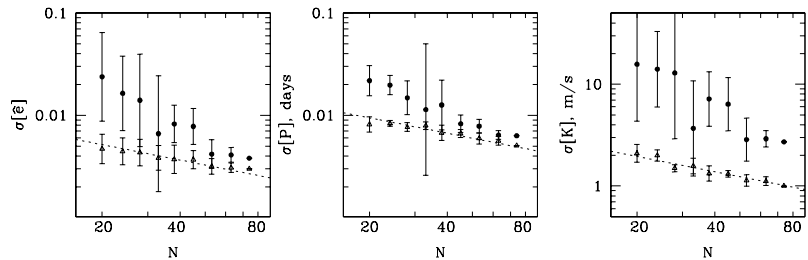}
\figcaption{Precision of different orbit parameters as a function of the size of the data set, for two values of input eccentricity (triangles for $e_{\rm in}=0$ and circles for $e_{\rm in}=0.6$) for the same system. Mock data sets are obtained by randomly throwing away some of the points from the complete set. Each number of points is sampled 6 times, and error bars correspond to the variance among these 6 realizations. The dotted line shows the best $N^{-1/2}$ fit to the precisions for $e_{\rm in}=0$. For $e_{\rm in}=0.6$ precisions, best-fit power-laws have slopes ranging from -1.5 to -1. \label{pic_shen}}
\end{figure}

To illustrate the effects of different $N$ in observing a single system, we take two mock radial velocity data sets, one with $\ein = 0$ and one with $\ein = 0.6$, for a high signal-to-noise system with a large number of observations (70 Vir) and decrease $N$ by deleting some observations at random. We then re-analyze the reduced data sets using our method and calculate orbit parameters and their precisions based on the MCMC output. The results are shown in Figure~\ref{pic_shen}. In this setup the precision of orbit parameters indeed improves as the number of observations $N$ increases: $\sigma$ decreases roughly as the expected $N^{-1/2}$ for $\ein = 0$ but for $\ein =0.6$ shows a scaling between $N^{-1.5}$ and $N^{-1.0}$. A possible explanation for this difference is that the improved phase coverage associated with larger $N$ improves orbit parameter determination more for a high-eccentricity system, where the cadence of observations near periastron is especially important \citep{endl06}, than for a low-eccentricity one.

\section{Correcting for eccentricity bias in radial velocity surveys}
\label{sec_true}

\subsection{Comparison with published orbit parameters}
\label{sec_compare}

In this section we apply our different eccentricity measures to real radial velocity data sets. Of the 91 B06 systems used to generate our mock radial velocity curves, we now consider the subsample of 65 systems which are well-fit by a single planet --- that is, those for which the root mean square residuals about the best-fit orbit solution for the system's largest planet are less than 15 m/s. Of these 65 systems, our Markov chains failed to converge for 14 Her. Since more recent work on 14 Her shows two planets with velocity amplitude ratio $K_c/K_b\simeq 0.3$ \citep{witt07}, we eliminated it from our subsample. Of the remaining 64 systems, HD 190360 also contains multiple planets. However, in this case our single-planet method finds the solution for the biggest planet, presumably because the second planet's relative contribution is smaller ($K_c/K_b\simeq 0.2$, \citealt{vogt05}).

\begin{figure}
\epsscale{0.7} 
%\plotone{picture_actual_paper1.eps}
\plotone{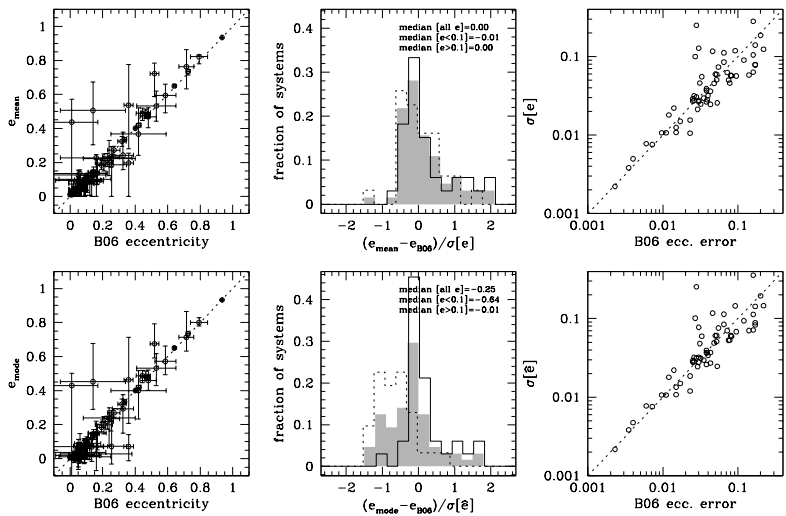}
\figcaption{Comparison between our eccentricity measurements and those from the B06 catalog for systems with a good one-planet fit. The top panels show the performance of the mean values of the posterior distribution $e_{\rm mean}$ and of their precisions. There is no systematic difference between these values and those from the B06 catalog (top middle; shaded grey for all eccentricities, dashed outline for those $<0.1$, and solid outline for those $>0.1$), so this is the measure we consider to be the closest to the published values. The bottom panels show the comparison between the published values and our preferred estimator $e_{\rm mode}$. This estimator returns values of eccentricity which are typically 0.25$\sigma$ lower than those in the published catalog (middle panel, shaded grey histogram). When we consider only planets with $e_{\rm mode}<0.1$ (dashed outline) this difference increases to 0.6$\sigma$, but for planets with $e_\mathrm{mode}>0.1$ alone this difference becomes negligible.\label{pic_butler}}
\end{figure}

For these 64 systems we compare our eccentricity estimators with published solutions, which are typically determined by fitting a Keplerian orbit to the radial velocity observations and minimizing $\chi^2$ in the relevant 7-parameter space. There are some minor differences between our analysis and that used for the B06 and other published solutions. In treating jitter, published solutions usually use observed stellar properties to fix $\sigma_J$ while we derive $\sigma_J$ from our MCMC output. Furthermore, the space in which our MCMC priors are flat --- namely, [$\ln P$, $\ln (1+K/K_o)$, $\omega$, $e$,   $M_0$, $C$, $\ln(1+\sigma_J/\sigma_{J,0})$] --- differs in its parametrization from the space over which published solutions typically minimize $\chi^2$.  Nevertheless, we find that our $e_{\rm mean}$, the mean of the marginalized posterior distribution for eccentricity, is statistically very similar to the B06 published eccentricities (see Figure~\ref{pic_butler}). The median difference between $e_{\rm mean}$ and the B06 catalog eccentricity is $<0.01\sigma[\hat{e}]$; this remains true when only planets with $e< 0.1$ are considered. We find that using $e_{\rm mode}$ results in eccentricity estimates systematically smaller than the published ones. The median difference between $e_{\rm mode}$ and the B06 values is $-0.25\sigma[\hat{e}]$; when we consider only $e_{\rm  mode}<0.1$ planets, it is $-0.6\sigma[\hat{e}]$.

\begin{figure}
\epsscale{0.7}
\plotone{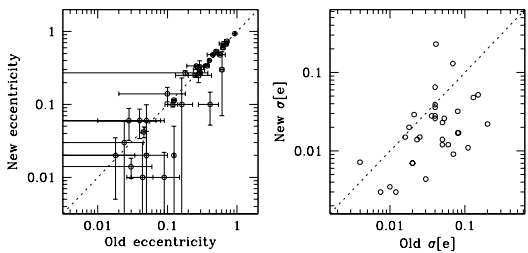}
\figcaption{The comparison between the discovery (`old') orbital eccentricites and their uncertainties and the most recent (`new') ones for the 34 systems from http://exoplanets.org for which six or more years separate the two epochs. The dotted lines show a one-to-one correspondence to guide the eye. As time went on, the effective signal-to-noise of observations increased, leading to decrease in eccentricity uncertainties and therefore in eccentricities themselves. \label{pic_decrease}}
\end{figure}

Our analysis in \S\ref{sec_corr} suggests that the bias in eccentricity should decrease with increasing quality (signal-to-noise) and number of observations. Using http://exoplanets.org, we selected all 34 planets in single-planet systems for which at least six years elapsed between the discovery announcement and the most recent published orbital solution. The comparison of eccentricity measurements and their uncertainties between the `old' (discovery) orbital solutions and the most recent `new' ones is presented in Figure~\ref{pic_decrease}. The uncertainties in eccentricity decreased as more and/or better observations were collected. Since eccentricity is a positively biased measure, as uncertainties decreased the values of eccentricity decreased as well. 

\subsection{Effects of choice of eccentricity estimator}
\label{sec_estimator}

In this section we examine the effect of the eccentricity estimators on the determination of the observed fraction of planets $f_{e_0}$ with eccentricities $\le e_0$ among the subsample of 64 B06 systems with good one-planet fits. We further exclude HD89307 for which Markov chains corresponding to the mock data sets did not converge due to poor sampling and few observations, yielding 63 systems. In cases where the $\sigma[\hat{e}]$ and hence the bias are comparable to the threshold $e_0$, we expect some nearly circular orbits to be misidentified as significantly eccentric. Where $\sigma[\hat{e}]\ll e_0$, the effect of bias should be negligible and the choice of eccentricity estimator is less critical. We report $f_{e_0}$ for multiple eccentricity estimators in Table~\ref{ecc_estimators}.

\begin{table}[h]
\begin{center}
\begin{tabular}{rcccc}
& eccentricity& & & \\
& estimator& $f_{0.02}$&  $f_{0.05}$&  $f_{0.1}$\\
\hline
all& B06& 0.06& 0.17& 0.38\\
& $e_\mathrm{mean}$& 0.06& 0.17& 0.38\\
& $e_\mathrm{mode}$& 0.23& 0.30& 0.49\\
& $\hat{e}_\mathrm{med}$& 0.16& 0.32& 0.49\\
\hline
$P>10$d& B06& 0.00& 0.10& 0.29\\
& $e_\mathrm{mean}$& 0.00& 0.10& 0.27\\
& $e_\mathrm{mode}$& 0.16& 0.23& 0.39\\
& $\hat{e}_\mathrm{med}$& 0.08& 0.25& 0.39\\
\end{tabular}
\caption{Summary of estimates of fractions of low-eccentricity
planets. \label{ecc_estimators}}
\end{center}
\end{table}

Although $e_{\rm mode}$ and $\hat{e}_{\rm med}$ are less biased than $e_{\rm mean}$ (Figure~\ref{pic_ecc_measures}), all three estimators must be biased for sufficiently circular orbits as they are positive definite. As an illustration, we consider the subsample of the $63\times 10=630$ simulations generated using the data for the 63 systems with input eccentricities $e_{\rm in}=0$ and $e_{\rm in}=0.05$. For the 315 simulations with $e_{\rm in}=0$, we measure $f_{0.05}=53\%$ using $e_{\rm mean}$ and $f_{0.05}=80\%$ using $e_{\rm mode}$, suggesting that this measure would still misidentify as eccentric a fifth of circular exoplanet orbits. For the simulations with $e_{\rm in}=0.05$ we measure $f_{0.05}=31\%$ using $e_{\rm mean}$ and $f_{0.05}=52\%$ using $e_{\rm mode}$. Using $e_{\rm mode}$ as an eccentricity estimator essentially eliminates the bias for $e_{\rm in}\ge 0.05$, but some bias remains for smaller eccentricities. 

\subsection{The fraction of planets on nearly circular orbits}
\label{sec_invert}

\begin{figure}
\epsscale{0.7} 
%\plotone{picture_fraction4.eps} 
\plotone{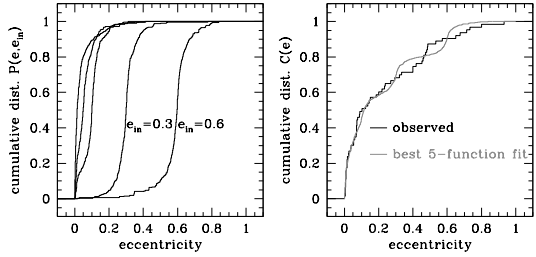} 
\figcaption{Left panel: cumulative distributions $P(e,e_{\rm in})$ (eq. \ref{eq_cumul}) determined from our simulations for all five input values $e_{\rm in}=0, 0.05, 0.1, 0.3, 0.6$ (increasing from left to right). The output measure of eccentricity is $e_{\rm mode}$ throughout this figure. Right panel: the cumulative distribution of eccentricities in the real extrasolar planetary systems (black) and its best approximation as a linear combination of the five functions $P(e,e_{\rm in})$ from our simulations (grey). The coefficients of this linear fit allow us to determine the intrinsic fraction of planets on nearly circular orbits: $f_{0.05}=0.33$. \label{pic_linfit}}
\end{figure}

In this section we estimate the true underlying eccentricity distribution, particularly near $e=0$ (Table 2). We focus on the 63 B06 systems which are well-fit by a single planet and for which mock data sets have convergent Markov chains. Let $p(e,e_{\rm in}) {\rm d}e$ be the probability that we observe a B06 system to have eccentricity between $e$ and $e+{\rm d}e$ if all the systems have true eccentricity $e_{\rm in}$. If all measurements were perfect,  $p(e,e_{\rm in})=\delta(e-e_{\rm in})$, but in practice this function is determined by the combination of eccentricity precisions in the B06 catalog. Our simulations yield $p(e,\ein)$ for $\ein=0, 0.05, 0.1, 0.3, 0.6$. If $D_{\rm in}(e_{\rm in})$ is the true eccentricity distribution in our 63-system subsample, then the observed distribution is
\begin{equation}
D_{\rm obs}(e)=\int_{0}^{1}D_{\rm in}(e_{\rm in}) p(e,e_{\rm in}) {\rm d}e_{\rm in}.\label{eq_dist}
\end{equation}
Integrating equation (\ref{eq_dist}) over $e$ from 0 to $e$ gives the observed
cumulative distribution
\begin{equation}
C_{\rm obs}(e)=\int_{0}^{1}D_{\rm in}(e_{\rm in}) P(e,e_{\rm in}) {\rm d}e_{\rm in},\label{eq_cumul}
\end{equation}
where $P(e,e_{\rm in})$ is the cumulative distribution corresponding to $p(e,e_{\rm in})$. In Figure~\ref{pic_linfit} (left), we show $P(e,e_{\rm in})$ for the five input values of eccentricity measured in our simulations.

As a first step we assume an underlying eccentricity distribution of the form $D_\mathrm{est}=A_0\delta(e-0) +A_{0.05}\delta(e-0.05)+A_{0.1}\delta(e-0.1) +A_{0.3}\delta(e-0.3) +A_{0.6}\delta(e-0.6)$ where the constants $A_i$ sum to unity. Estimating the underlying eccentricity distribution then amounts to finding coefficients $A_i$ such as $C_{\rm est}(e)=\sum_i A_i P(e,e_{{\rm in,}i})$ best represents the observed cumulative distribution $C_{\rm obs}(e)$. We find the best-fit $A_i$ by minimizing the Kolmogorov-Smirnov (KS) statistic between $C_{\rm obs}(e)$ and $C_{\rm est}(e)$ (Figure~\ref{pic_linfit}, right). Using $e_{\rm mode}$ as the estimator, we obtain $A_0=0.26$, $A_{0.05}=0.15$, $A_{0.1}=0.19$, $A_{0.3}=0.20$, $A_{0.6}=0.21$, putting 33\% of the 63 planets on orbits with $e\leq 0.05$. In principle, this method can be used with any eccentricity estimator, as long as the function $P(e,\ein)$ represents the cumulative distribution for the same estimator. The derived $f_{0.05}$ varies in the range $21\%-45\%$ depending on the estimator, with 33\% being the median value.

\begin{figure}
\begin{center}
\includegraphics*[scale=.4]{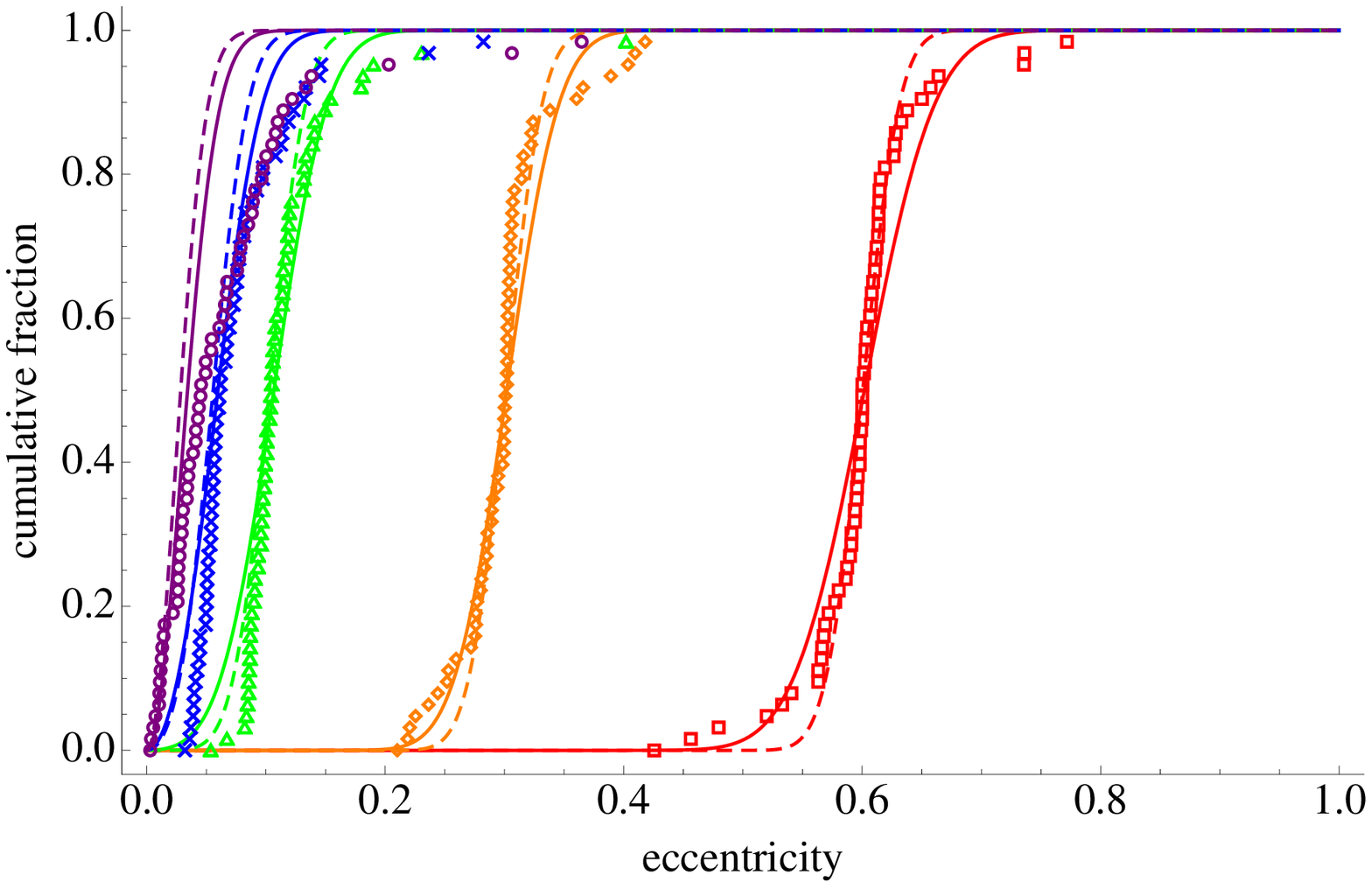} 
\includegraphics*[scale=.4]{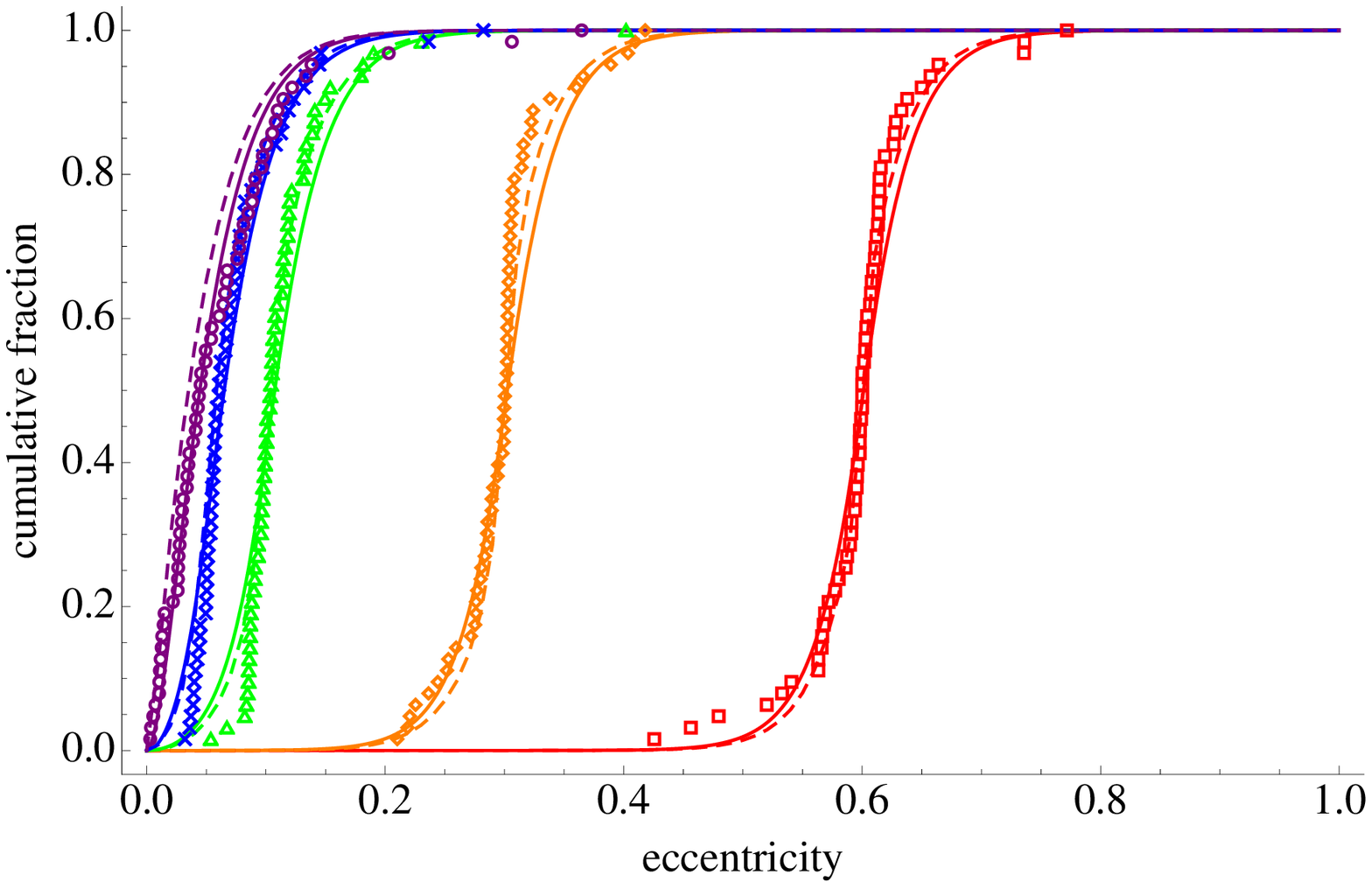}
\end{center}
\figcaption{Comparison between $P(e,\ein)$ derived from our MCMC simulations, with $e_{\rm mean}$ as the eccentricity measure, and the analytic approximations from \S\ref{sec_invert} (the points and curves are, from left to right, for $\ein =0$, $\ein =0.05$, $\ein =0.1$, $\ein =0.3$ and $\ein =0.6$, or purple, blue, green, orange and red, correspondingly, in the on-line version of the paper). The top panel shows Gaussian models (eq. \ref{eq_gaussfit}). For each $P(e,\ein)$, we try two different values of $\Sigma$. For small eccentricities and data sets with many observations evenly distributed over orbit phase, we expect $\Sigma\simeq{\rm median}(\sqrt{2}\sigma_{\rm obs}/K\sqrt{N})$, shown with solid lines. We also try $\Sigma={\rm median}(\sigma[\hat{e}])$, shown with dashed lines. While the former produces nominally better fits than the latter, both sets poorly describe the tails of the MCMC result distributions, especially for $\ein =0$ and $\ein =0.05$. The bottom panel shows single exponential models with the fitted $\Sigma=0.0302$ (dashed lines) and double exponential models with $\Sigma_1=0.0305$, $\Sigma_2=0.00874-0.00719\ein$, $B=3.46+20.9\ein$ (solid lines). Exponential models are significantly better than the Gaussian for reproducing the tails of the MCMC-derived distributions. The double exponential model is better than the single exponential one for all $\ein $ values except $\ein=0$. \label{efuncfits}}
\end{figure}

In reality, we expect a continuous underlying eccentricity distribution. Because it is impractical to calculate $P(e,e_{\rm in})$ via Monte Carlo simulations for a large number of input eccentricities, our method is to adopt an analytical form for $P(e,e_{\rm in})$ which agrees with the functions derived for the five values of $\ein$ in mock systems and may be generalized for other values of $\ein$. We use $e_{\rm mean}$ as our eccentricity estimator for this method, as this is the estimator closest to the one used in real radial velocity surveys.  If all systems in the B06 catalog had the same eccentricity precision and if the observed values of $e\cos\omega$ and $e\sin\omega$ followed Gaussian distributions with dispersion $\Sigma$ about their true values \citep{shen08}, then
\begin{equation}
p(e,e_{\rm in}) = \frac{e \exp\left(-\frac{e^2}{2\Sigma^2}\right) I_0\left(\frac{e e_{\rm in}}{\Sigma^2}\right)}
{\int_0^1 {\rm d}e' e' \exp\left(-\frac{(e')^2}{2\Sigma^2}\right) I_0\left(\frac{e' e_{\rm in}}{\Sigma^2}\right)}
\label{eq_gaussfit}
\end{equation}
and $P(e,e_{\rm in}) = \int_0^e p(e',e_{\rm in}) {\rm d} e'$. As we show in Figure~\ref{efuncfits}, such distribution does not accurately reproduce $P(e,e_{\rm in})$ from our simulations. This occurs because the catalog combines systems with a range of eccentricity precisions. 

Much better fits to $P(e,e_{\rm in})$ from simulations are obtained by assuming that $e \cos\omega$ and $e \sin\omega$ follow an exponential distribution in $(-|e-\ein |)$ or a sum of two exponential distributions. For our single exponential model, we take
\begin{equation}
p(e,\ein ) = \frac{e\int_0^{2\pi}{\rm d}\omega \exp\left(-\frac{\sqrt{e^2+\ein ^2-2e \ein \cos\omega}}{\Sigma/|\ln 0.32|}\right)}{\int_0^1 {\rm d}e' \, e'\int_0^{2\pi}{\rm d}\omega \exp\left(-\frac{\sqrt{(e')^2+\ein ^2 -2e' \ein \cos\omega}} {\Sigma/|\ln 0.32|}\right)}
\label{eq_single}
\end{equation}
where the factors of $|\ln0.32|$ are included so that $\Sigma$ corresponds to the 68\% confidence interval in $h$ or $k$. We emphasize that $\Sigma$ in this expression is not an eccentricity uncertainty for any specific data set, but a fitting parameter for the entire ensemble of systems. We choose $\Sigma=0.0302$ by minimizing the sum of the KS statistics between each of the MCMC result distributions for the five $\ein $ and its corresponding model $p(e,\ein )$. Similarly, for our double exponential model we take
\begin{equation}
p(e,\ein )
 = \frac{e\int_0^{2\pi}d\omega\left[
            \exp\left(-\frac{\sqrt{e^2+\ein ^2
                             -2e \ein \cos\omega}}
                            {\Sigma_1/|\ln 0.32|}\right)
            + B\exp\left(-\frac{\sqrt{e^2+\ein ^2
                                -2e \ein \cos\omega}}
                               {\Sigma_2/|\ln 0.32|}\right)\right]}
        {\int_0^1 {\rm d}e' \, e'\int_0^{2\pi}d\omega\left[
                      \exp\left(-\frac{\sqrt{(e')^2+\ein ^2
                                             -2e' \ein \cos\omega}}
                                      {\Sigma_1/|\ln 0.32|}\right)
                      + B\exp\left(-\frac{\sqrt{(e')^2+\ein ^2
                                                -2e' \ein \cos\omega}}
                                         {\Sigma_2/|\ln 0.32|}\right)\right]}
\;\;\; ,
\label{eq_double}
\end{equation}
with three parameters --- the widths $\Sigma_1$, $\Sigma_2$ and the amplitude $B$ --- which vary linearly with $\ein $. The double exponential models are somewhat better fits than the single exponential models for all $\ein $ except $\ein =0$ (Figure~\ref{efuncfits}). Models (\ref{eq_single})-(\ref{eq_double}) yield similar cumulative eccentricity distributions and $f_{0.02}$, $f_{0.05}$ values so below we discuss the single exponential model. 

\begin{figure}
\begin{center}
%\PSbox{underlyingall.ps hscale=65 vscale=65 voffset=-30}{408pt}{180pt}
%\PSbox{underlyinglong.ps hscale=65 vscale=65 voffset=-30}{408pt}{220pt}
%\PSbox{fig11a.ps hscale=30 vscale=30 voffset=-30}{408pt}{180pt}
%\PSbox{fig11b.ps hscale=30 vscale=30 voffset=-30}{408pt}{220pt}
\includegraphics*[scale=.4]{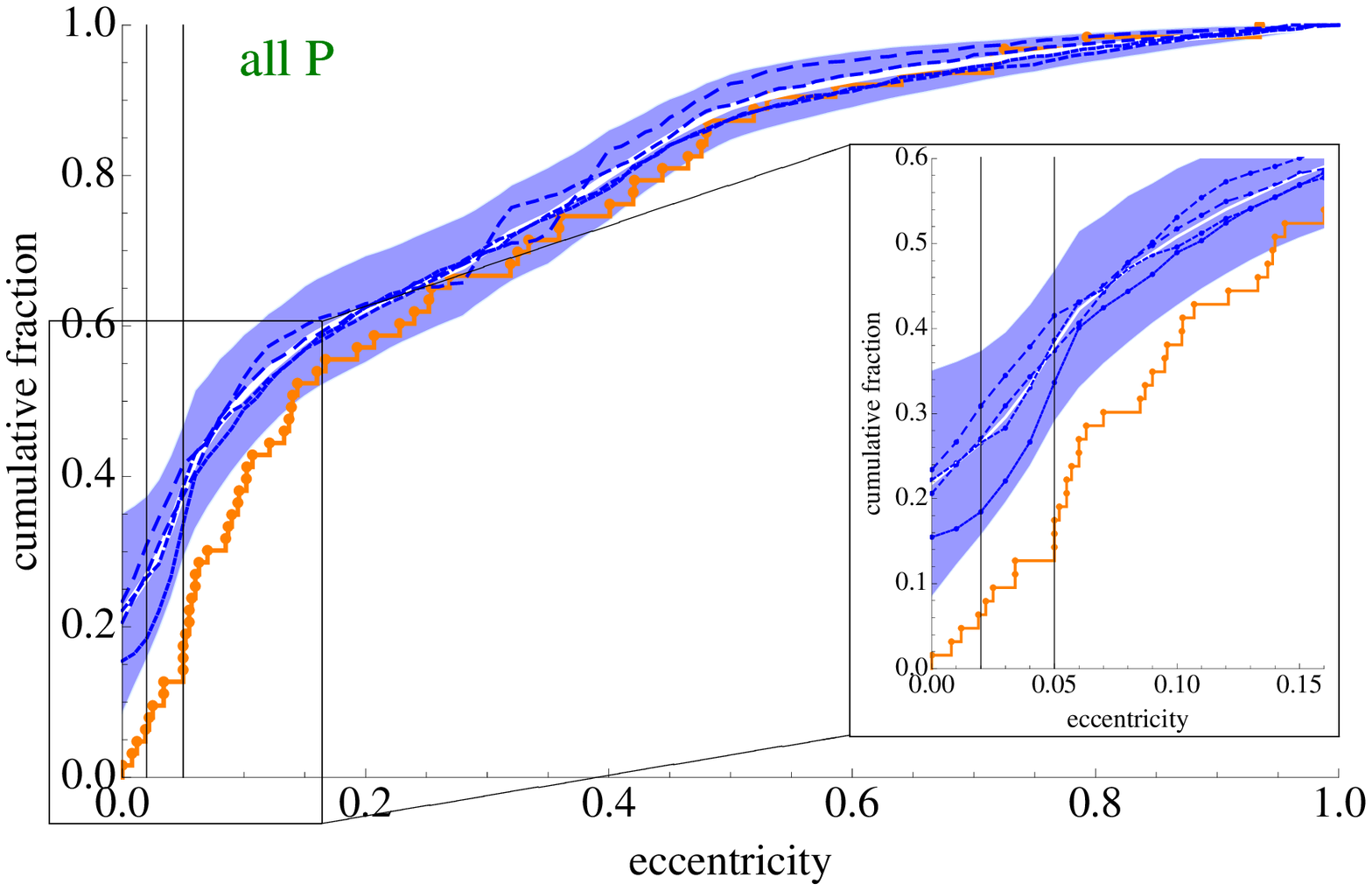} 
\includegraphics*[scale=.4]{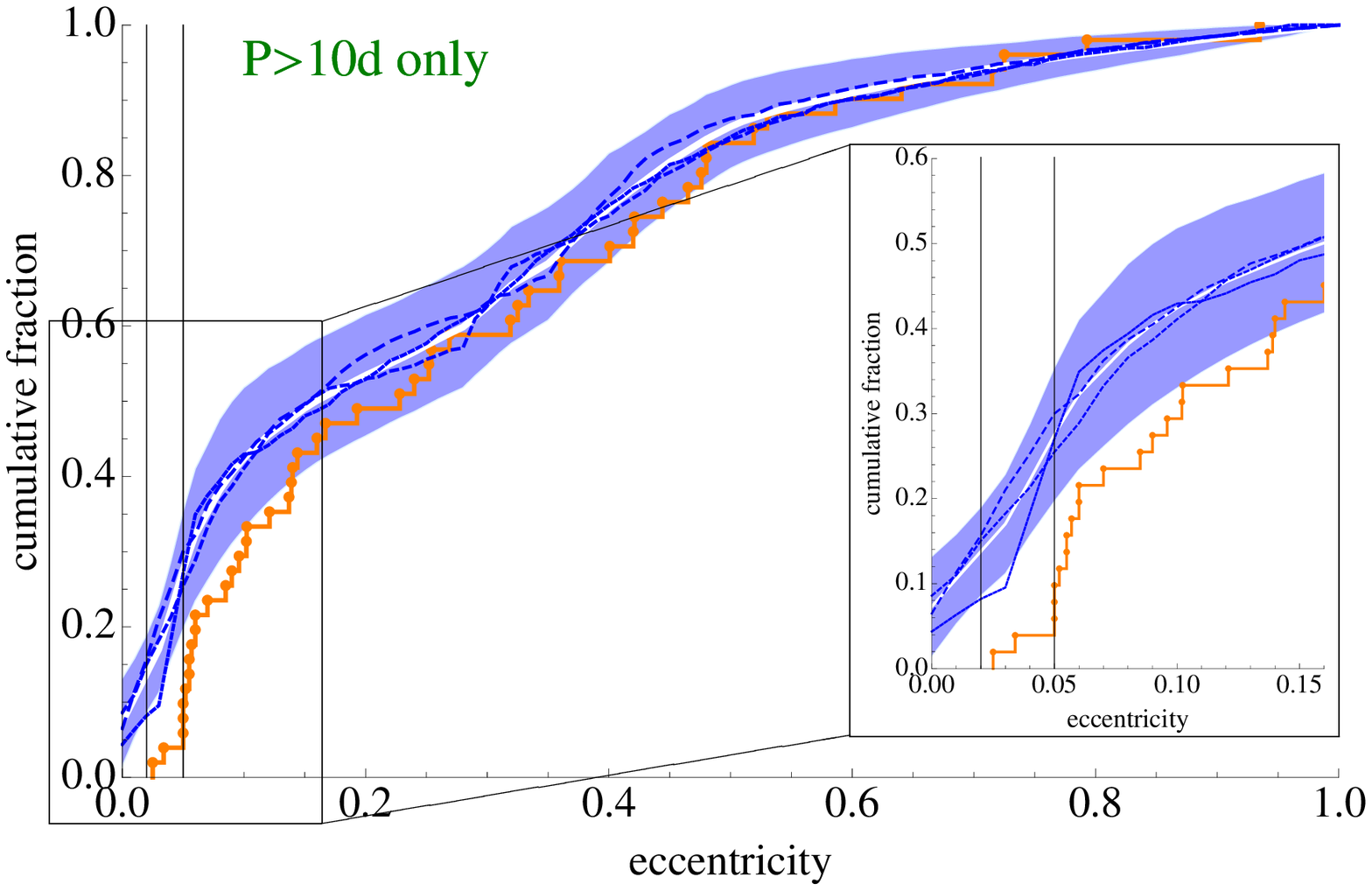}
\end{center}
\figcaption{Estimates of the underlying cumulative eccentricity distribution for 63 B06 systems (top plot) and for the 51-system subset of those with period longer than 10 days (bottom plot). In both plots the inset panel provides a larger view of the lower left-hand corner of the main plot. Results of a bootstrap analysis using a single exponential model for $p(e,e_\mathrm{in})$ with $e_\mathrm{in}$ grid spacings 0.02, 0.03, 0.04, 0.05 in the top plot and 0.03, 0.04, 0.05 in the bottom plot appear as dashed lines; longer dashes correspond to larger grid spacings. The white line shows the average of the results obtained using the different grid spacings, and the shading gives the $1\sigma$ uncertainty obtained by averaging the bootstrap standard deviations obtained using the four grid spacings. As the top plot indicates, the implied underlying fractions of single-planet systems with eccentricities below 0.02, 0.05 --- respectively $f_{0.02}=0.27\pm0.12$, $f_{0.05}=0.38\pm0.09$ --- are larger than the $f_{0.02}=0.06$, $f_{0.05}=0.17$ of the published eccentricities (orange) by $1.9\sigma$ and $2.3\sigma$. For the systems with $P>10d$ the bottom plot indicates underlying values $f_{0.02}=0.13\pm 0.05$, $f_{0.05}=0.28\pm 0.08$, which are larger than the $f_{0.02}=0$, $f_{0.05}=0.10$ by $2.6$ and $2.3$ standard deviations, respectively. We include vertical lines at $e=0.02,0.05$ to guide the eye.\label{underlying}}
\end{figure}

We discretize the intrinsic eccentricity distribution as
\begin{equation}
D_\mathrm{est}(e)=\sum_i A_i \delta(e-e_{{\rm in},i}) \label{eq_grid}
\end{equation}
and fit for $A_i$ as in our first example of five input eccentricities. For the grid of $e_i$ used in $D_\mathrm{est}$ we try eccentricity spacings 0.02, 0.03, 0.04, 0.05 and perform a bootstrap analysis on the inversion using each of the four $\ein$ grids. The underlying eccentricity distributions inferred using the four grids are close to one another (Figure~\ref{underlying}) and all are well above the observed eccentricity distribution for $e\le 0.1$. The resulting fractions of planets on nearly circular orbits are $f_{0.02}=0.27\pm 0.11$ and $f_{0.05}=0.38\pm 0.09$, much higher than the B06 values of 0.06 and 0.17, respectively. We apply the same analysis to the 51 systems with $P>10d$ and find underlying values $f_{0.02}=0.13\pm 0.05$, $f_{0.05}=0.28\pm 0.08$, much larger than the B06 values of $f_{0.02}=0$ and $f_{0.05}=0.10$. We can likewise apply this analysis to the set of all 117 B06 single-planet systems if we assume that it is sufficiently statistically similar to the subset of the 63 planets in our simulations that the same $P(e,\ein)$ can be used in both cases. This yields underlying values $f_{0.02}=0.30\pm 0.08$, $f_{0.05}=0.32\pm 0.06$, whereas on the basis of the published eccentricities we would calculate $f_{0.02}=0.06$ and $f_{0.05}=0.18$. 

Another method to obtain the intrinsic eccentricity distribution is to directly solve for $D_{\rm in}(\ein)$ in equation (\ref{eq_dist}) with functions $p(e,\ein)$ from equation (\ref{eq_single}) using the iterative deconvolution procedure of \citet{lucy74}. Specifically, given the 0th guess for the intrinsic eccentricity distribution $D_{\rm in}^0(\ein)$ (e.g., a flat distribution), the subsequent iterations are obtained using
\begin{equation}
D_{\rm in}^{r+1}(\ein)=\frac{D_{\rm in}^r(\ein)}{N}\sum_{i=1}^N\frac{p(e_i,\ein)}{\int_0^1 D_{\rm in}^r(\ein)p(e_i,\ein){\rm d}\ein}.
\label{eq_lucy_1}
\end{equation}
This method yields $f_{0.05}=24\%$, and the result of the deconvolution is illustrated in Figure \ref{pic_lucy}a. 

Our model functions $P(e,e_{\rm in})$ from equation (\ref{eq_single}) deviate slightly from the results of the simulations. In particular, for $\ein=0.05$ and $\ein=0.1$, the model functions overpredict $P(e,\ein)$ by $\la 0.1$ (Figure \ref{efuncfits}, bottom). To estimate the uncertainties introduced into $f_{e_0}$ by using the model approximation, we expand the integrand of equation (\ref{eq_cumul}) around the model functions, $D^{\rm actual}=D^{\rm model}+\Delta D$ and $P^{\rm actual}=P^{\rm model}+\Delta P$, retain the first order in corrections $\Delta D$ and $\Delta P$ and use the constraint $C_{\rm obs}(e)=\int_{0}^{1}D^{\rm model}_{\rm in}(e_{\rm in}) P^{\rm model}(e,e_{\rm in}) {\rm d}e_{\rm in}$. This allows us to relate the correction $\Delta D$ to the known functions $P^{\rm model}$, $D^{\rm model}$ and $\Delta P$ (the latter can be estimated from Figure \ref{efuncfits}). This procedure suggests that $f_{0.1}$ derived using model functions is overestimated by about 0.03, well within the uncertainties of the deconvolution procedure. We note that the 5-functions method described in the beginning of this section is not affected by this systematic error and therefore provides an independent check. 

\begin{table}[h]
\begin{center}
\begin{tabular}{rcccc}
& eccentricity& & & \\
method & estimator& $f_{0.02}$&  $f_{0.05}$&  $f_{0.1}$\\
\hline
5-functions & $e_{\rm mean}$ & 0.13& 0.21& 0.45\\
5-functions & $e_{\rm med}$ & 0.29& 0.31& 0.47\\
5-functions & $e_{\rm mode}$ & 0.26& 0.33& 0.50\\
grid & $e_{\rm mean}$ & 0.27$\pm$0.11 & 0.38$\pm$0.09 & 0.51$\pm$0.08 \\
Lucy deconvolution (eq. \ref{eq_single}, \ref{eq_lucy_1}) & $e_\mathrm{mean}$& 0.18& 0.24& 0.49\\
\hline
$P>10$d, grid & $e_{\rm mean}$ & 0.13$\pm$0.05 & 0.28$\pm$0.08 & 0.42$\pm$0.09 \\
\end{tabular}
\caption{Estimates of the true underlying fraction of planets on nearly circular orbits using different methods\label{tab_models}}
\end{center}
\end{table}

\section{Summary}
\label{sec_summary}

In this work, we constructed a catalog of mock radial velocity data for 2275 artificial single-planet systems with eccentricities 0, 0.05, 0.1, 0.3, and 0.6 and all other orbital parameters drawn from the real radial velocity data sets. We analyzed these data using MCMC simulations and compared the input and extracted orbital parameters in order to study potential biases introduced into the population of known radial velocity single-planet systems by the orbit extraction process.

We paid particular attention to eccentricity biases because of its significance for testing planet formation models. Eccentricity is positive definite, and therefore a measurement bias is present, especially for planets on small-eccentricity orbits. The mode of the marginalized posterior one-dimensional eccentricity distributions output by MCMC simulations was the least biased of the eccentricity estimators we considered . The mode outperformed the mean $e_\mathrm{mean}$ and median $e_\mathrm{median}$ of the posterior eccentricity distribution as well as estimators based on mean or median values of $h=e\cos\omega$ and $k=e\sin\omega$. Since our $e_\mathrm{mean}$ closely reproduces eccentricities derived using the standard $\chi^2$-minimization methods, we suggest that the eccentricities derived and reported for planets with intrinsic eccentricities $\leq 0.05$ are typically biased high by $1\sigma-2\sigma$, while those for planets with intrinsic eccentricities $\geq 0.1$ are not significantly biased. We recommend $e_{\rm mode}$, the mode of the marginalized posterior distribution for eccentricity, as the preferred eccentricity estimator in observational studies of exoplanet population statistics. This requires minimal effort as many radial velocity exoplanet surveys already use MCMC analysis for orbital parameter estimation.

The study most closely related to ours is that of \citet{shen08}. These authors isolate the dependence of the quality of individual derived orbit parameters on the number of observations and on the signal-to-noise ratio by varying these parameters separately in the radial-velocity data for one single-planet system and extracting orbits via $\chi^2$ minimization. Their input orbit parameters differ somewhat from ours; in particular, they include many systems with much lower weighted signal-to-noise than ours and they do not consider the maximum phase gap or number of periods covered. Their findings that eccentricity bias preferentially affects low-eccentricity systems and that eccentricity errors decrease strongly with increasing weighted signal-to-noise are qualitatively consistent with ours. Our approach emphasizes the role of a realistic ensemble of planetary systems in shaping the relations between input data set quality and output orbit quality, particularly eccentricity bias. Since we draw our systems from a real survey, this allows us to estimate the underlying fraction of nearly circular exoplanet orbits. 

\begin{figure}
\epsscale{0.7} 
%\plotone{picture_decompose_distribution_int3.eps}
\plotone{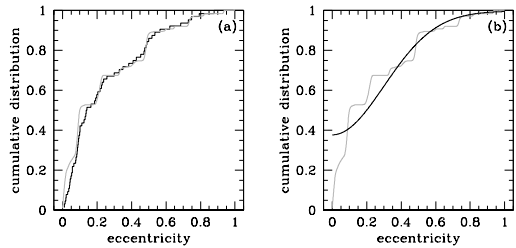}
\figcaption{Left: Observed cumulative eccentricity distributions in black and the Lucy-deconvolved eccentricity distribution in grey. The eccentricity estimator is $e_{\rm mean}$, and \citet{lucy74} deconvolution is applied as in equation (\ref{eq_lucy_1}) with functions $p$ given by equation (\ref{eq_single}). Right: The deconvolved distribution (grey) is represented as a sum of two populations: a population of planets on circular orbits and another population of planets with an eccentricity distribution ${\rm d}N \propto e \exp(-\frac{1}{2}(e/0.3)^2){\rm d} e$. The solid black line is the best linear fit, with 38\% of all systems attributed to the former component, meant to represent systems which are dynamically inactive or have been circularized, and 62\% to the latter, representing dynamically active systems resulting from planet-planet scattering. \label{pic_lucy}}
\end{figure}

Using several methods, we estimate the true underlying fraction of planets on nearly circular orbits among B06 single-planet systems. We find that the fractions of planets with eccentricities below $0.02$ and below $0.05$ are respectively $f_{0.02}=0.27\pm 0.11$ and $f_{0.05}=0.38\pm 0.09$ --- significantly higher than $f_{0.02}=0.06$, $f_{0.05}=0.17$ computed using the B06 published eccentricities. When we exclude systems with periods less than 10 days, we find values $f_{0.02}=0.13\pm 0.05$, $f_{0.05}=0.28\pm 0.08$, again significantly larger than the $f_{0.02}=0$, $f_{0.05}=0.10$ of the B06 published eccentricities. This suggests that low eccentricities like those seen among major planets in our solar system may not be as unusual among radial velocity exoplanets as has previously been believed. In particular, the eccentricity distribution (corrected for biases) is not well matched to the eccentricity distribution of dynamically active systems that went through a phase of planet-planet scattering, ${\rm d}N\propto e \exp(-\frac{1}{2}(e/0.3)^2){\rm d}e$ \citep{juri08}, which would result in almost no planets on nearly circular orbits. In Figure \ref{pic_lucy}b, we represent the instrinsic (de-biased) eccentricity distribution as a linear combination of a population of planets on circular orbits (38\% of systems) and a population of dynamically active systems described by the \citet{juri08} distribution (62\%). Therefore, a large fraction of all planets may have avoided a phase of planet-planet scattering, for example because they were formed in systems with few very massive planets. Alternatively, these systems may have been dynamically active, but the eccentricities may have been damped by some mechanism (e.g., by the residual disk material, \citealt{raym09, mats10}). 

While we found no overall bias among periods and velocity amplitudes extracted from our mock planet catalog, the errors we derived for our periods, velocity amplitudes, and eccentricities suggest rough guidelines for the quality of extracted orbit parameters one might reasonably expect from real radial velocity data sets. Specifically, an eccentricity precision $\leq 0.05$ typically requires weighted signal-to-noise $\geq 40$; a normalized period precision of 1\% is achieved in only two-three orbital periods; and a velocity amplitude precision of 3~m/s typically requires a maximum phase gap $\leq 0.3$ period for low- to moderate-eccentricity systems and $\leq 0.15$ period for moderate- to high-eccentricity systems. These guidelines correspond to the median relationships seen in our data. 

Limitations of our study include our restricting our mock catalog to single-planet systems with fewer than 90 observations. Thus, our results exclude the best-studied systems. Historically, there is a tendency for reported eccentricities to decrease with more observations (\citealt{butl06} and \S\ref{sec_compare}), sometimes due to the discovery of additional planets. Nevertheless, we expect that our main findings can be generalized to other radial velocity surveys. Due to competition for observing time, few systems are followed up simply to achieve higher precision in parameter determination. Thus eccentricity bias remains significant at low eccentricity, and simply increasing the sample size of known planets is insufficient to ensure that the observed eccentricity distribution approaches the underlying one.

Our results also exclude the $\sim$30\% of radial-velocity planets found in multiple-planet systems. Such planets have an eccentricity distribution similar to that of single-planet systems \citep{wright09}, so eccentricity bias of the kind described in this work is likely an important consideration in the population statistics of multiple planet systems as well. 

\acknowledgments
N.L.Z. was supported by the {\it Spitzer} Space Telescope Fellowship provided by NASA through a contract issued by the Jet Propulsion Laboratory, California Institute of Technology; by the John N. Bahcall Fellowship at the IAS; and by the NSF grant AST-0807444. M.P. was supported by AMIAS and Taplin memberships at the IAS. E.B.F. was supported by NASA Origins of Solar Systems grant NNX09AB35G and the University of Florida. M.P. and E.B.F. were supported in part by the NSF grant PHY05-51164, by Kavli Institute for Theoretical Physics at University of California, Santa Barbara, and by the Aspen Center for Physics. The authors would like to thank Mario Juric and Scott Tremaine for discussions and the referee for the comments on the manuscript.

\begin{thebibliography}{}
\bibitem[Adams \& Laughlin(2003)]{adam03}Adams, F.C. \& Laughlin, G. 2003, Icarus, 163, 290
\bibitem[Balan \& Lahav(2009)]{bala09}Balan, S.T. \& Lahav, O. 2009, \mnras, 394, 1936
\bibitem[Batygin et al.(2009)]{baty09}Batygin, K., Laughlin, G., Meschiari, S., Rivera, E., Vogt, S., \& Butler, P. 2009, \apj, 699, 23
\bibitem[Butler et al.(2006)]{butl06}Butler et al. 2006, \aj, 646, 505 (B06)
\bibitem[Chatterjee et al.(2008)]{chat08}Chatterjee, S., Ford, E.B., Matsumura, S., \& Rasio, F.A. 2008, \apj, 686, 580
\bibitem[Cumming(2004)]{cumm04}Cumming, A. 2004, MNRAS, 354, 1165
\bibitem[Endl et al.(2006)]{endl06}Endl, M., Cochran, W.D., Wittenmyer, R.A., \& Hatzes, A.P. 2006, \aj, 131, 3131
\bibitem[Fabrycky \& Tremaine(2007)]{fabr07}Fabrycky, D. \& Tremaine, S. 1007, \apj, 669, 1298
\bibitem[Ford(2005)]{ford05}Ford, E.B. 2005, \aj, 129, 1706
\bibitem[Ford(2006)]{ford06a}Ford, E.B. 2006, \apj, 642, 505
\bibitem[Ford(2008)]{ford08b}Ford, E.B., 2008, \apj, 135, 1008
\bibitem[Ford \& Gregory(2007)]{ford07}Ford, E.B. \& Gregory, P.C. 2007, ASP Conf. Ser., 371, 189
\bibitem[Ford \& Rasio(2006)]{ford06b}Ford, E.B. \& Rasio, F.A. 2006, \apj, 638, L45
\bibitem[Ford \& Rasio(2008)]{ford08a}Ford, E.B. \& Rasio, F.A. 2008, \apj, 686, 621
\bibitem[Ford et al.(2001)]{ford01}Ford, E.B., Havlickova, M., \& Rasio, F.A. 2001, Icarus, 150, 303
\bibitem[Gregory(2005)]{greg05}Gregory, P.C. 2005, \apj, 631, 1198
\bibitem[Gregory(2007)]{greg07}Gregory, P.C. 2007, \mnras, 374, 1321
\bibitem[Holman et al.(1997)]{holm97}Holman, M., Touma, J., \& Tremaine, S. 1997, Nature, 386, 254
\bibitem[Juri\'{c} \& Tremaine(2008)]{juri08}Juri\'c, M. \& Tremaine, S. 2008, \apj, 686, 603
\bibitem[Laughlin \& Adams(1998)]{laug98}Laughlin, G. \& Adams, F.C. 1998, \apj, 508, L171
\bibitem[Lucy(1974)]{lucy74}Lucy, L.B. 1974, \aj, 79, 745
\bibitem[Lucy \& Sweeney(1971)]{lucy71}Lucy, L.B. \& Sweeney, M.A. 1971, AJ, 76, 544
\bibitem[Malmberg \& Davies(2009)]{malm09}Malmberg, D. \& Davies, M.B. 2009, \mnras, 394, L26
\bibitem[Malmberg et al.(2007)]{malm07}Malmberg, D., de Angeli, F., Davies, M.B., Church, R.P., Mackey, D., \& Wilkinson, M.I. 2007, \mnras, 378, 1207
\bibitem[Marzari \& Weidenschilling(2002)]{marz02}Marzari, F. \& Weidenschilling, S.J. 2002, Icarus, 156, 570
\bibitem[Marzari et al.(2005)]{marz05}Marzari, F., Weidenschilling, S.J., Barbieri, M., \& Granata, V. 2005, \apj, 618, 502
\bibitem[Matsumura et al.(2008)]{mats08}Matsumura, S., Takeda, G., \& Rasio, F.A. 2008, \apj, 686, L29
\bibitem[Matsumura et al.(2010)]{mats10}Matsumura, S., Thommes, E.W., Chatterjee, S., Rasio, F.A. 2010, \apj, 714, 194
\bibitem[Murray \& Dermott(1999)]{murr99}Murray, C.D. \& Dermott, S.F. 1999, Solar System Dynamics (Cambridge University Press, Cambridge, New York)
\bibitem[Nagasawa et al.(2008)]{naga08}Nagasawa, M., Ida, S., \& Bessho, T. 2008, \apj, 498
\bibitem[O'Toole et al.(2008)]{otoo08}O'Toole, S.J., Tinney, C.G., Jones, H.R.A., Butler, R.P., Marcy, G.W., Carter, B., \& Bailey, J. 2009, MNRAS, 392, 641
\bibitem[Rasio \& Ford(1996)]{rasi96}Rasio, F.A. \& Ford, E.B. 1996, Science, 274, 954
\bibitem[Raymond et al.(2009)]{raym09}Raymond, S.N., Armitage, P.J., Gorelick, N. 2009, \apjl, 699, L88
\bibitem[Shen \& Turner(2008)]{shen08}Shen, Y. \& Turner, E.L. 2008, \apj, 685, 553
\bibitem[Tremaine \& Zakamska(2004)]{trem04}Tremaine, S. \& Zakamska, N.L., 2004, AIP Conf. Proc., 713, 243
\bibitem[Takeda \& Rasio(2005)]{take05}Takeda, G. \& Rasio, F.A. 1005, \apj, 627, 1001
\bibitem[Vogt et al.(2005)]{vogt05}Vogt, S.S., Butler, R.P., Marcy, G.W., Fischer, D.A., Henry, G.W., Laughlin, G., Wright, J.T., \& Johnson, J.A. 2005, \apj, 632, 638
\bibitem[Weidenschilling \& Marzari(1996)]{weid96}Weidenschilling, S.J. \& Marzari, F., 1996, Nature, 384, 619
\bibitem[Wittenmyer et al.(2007)]{witt07}Wittenmyer, R.A., Endl, M., \& Cochran, W.D. 2007, \apj, 654, 625
\bibitem[Wright(2005)]{wrig05}Wright, J.T. 2005, PASP, 117, 657
\bibitem[Wright(2009)]{wright09}Wright, J.T. 2009, arXiv:0909.0957 (conf. proc., submitted)
\bibitem[Wright \& Howard(2009)]{wrig09}Wright, J.T., \& Howard, A.W. 2009, ApJS, 182, 205
\bibitem[Zakamska \& Tremaine(2004)]{zaka04}Zakamska, N.L. \& Tremaine, S. 2004, \aj, 128, 869
\end{thebibliography}
\end{document}